\documentclass[aps,prd,superscriptaddress,eqsecnum,amsfonts,showpacs,epsfig]{revtex4}
\usepackage{epsfig}

\newcommand{\be}{\begin{equation}}
\newcommand{\ee}{\end{equation}}
\newcommand{\bea}{\begin{eqnarray}}
\newcommand{\eea}{\end{eqnarray}}

\newcommand{\nn}{\nonumber}

\begin{document}


\title{The effective charm mass from the excited charmonium leptonic decays}

\author{V. \v{S}auli}

\email{sauli@ujf.cas.cz}
\affiliation{Department of Theoretical Physics, Institute of Nuclear Physics Rez near Prague, CAS, Czech Republic  }

\begin{abstract}
We use the covariant four-dimensional Bethe-Salpeter (BS) equation to determine the effective charm quark mass.
The scale dependence of the effective charm quark mass is determined using experimentally known spin-one charmonium spectra and leptonic decay constants. The infrared finite, massive-like effective QCD running charge is used to solve the Bethe-Salpeter equation to achieve this. The obtained effective charm quark mass values run from 1.1 GeV for the $J/Psi$  meson to 1.5 GeV  for its higher  radial excitations.  These values are substantially smaller than those extracted in the perturbative MS renormalization scheme, but they agree with the semi-perturbative solution of the quark Schwinger-Dyson equation in the momentum subtraction scheme. 

Within the interaction entirely governed by QCD running coupling,  the sliding scale charm quark mass  
is crucial for the correct determination of leptonic decays.
Notably, for leptonic decay constants of excited states, this is the first time the theory based on Bethe-Salpeter equation has met the precision of the experimental data.     
\end{abstract}
\pacs{11.10.St, 11.15.Tk}
\maketitle

\section{Introduction}

The study of the heavy quarkonium properties was initiated long time ago, using potential models. A large variety of them exists in the literature and there still remains the question of their connection with the QCD  parameters. In practice, the effective nonrelativistic theory of QCD \cite{BPSV2005,BPSV2000} made significant advances. However, it is still unclear what the original object was that was taken to this nonrelativistic limit.    The BS equation was used to derive the known QED formula \cite{KK1952}, which  shortly after QCD discovery,  was  adapted for the evaluation of radiative corrections to the leptonic quarkonia decays \cite{BGKK1975}. Six years later, it was elegantly  rederived in \cite{BT1981}.  As indicated  there, the nonrelativistic limit could emerge from the QCD Green's function, which are the solution of  the Bethe-Salpeter (BS) equation for mesons. 
To make a profit from lattice calculation the nonrelativistic QCD was further matched with Wilson loop operators. This could be   perhaps useful when interpreting long range part potentails. The short-range part, as advocated in \cite{BT1981}, is governed by asymptotic freedom and  described by the running strong coupling. There is no nig experience of how to use the BSE for quarkonia. While quarkonia should be an excelent laboratory for BSE studies, this approach is mostly abandoned  in the literature. For a recent  development of heavy quarkonia in various
contexts, see \cite{EGMR2008,VOL2008,BRA2011,PPR2012,LAN2020,Chap2022,EMPP2026}. In this paper, we try to take another look at this issue again and solve the BSE using at least two  methods.  We find that it is difficult to reach the desired non-relativistic limit if we are not sure about the effect of the running quark mass.  We will use a quite  primitive method, which allows us to demonstrate that the running quark  mass can supply the confining part of the nonrelativistic potential. The later  is not created  in strict mathematical sense, but  it supplies purely relativistc effect of the running mass very effectively at least in the spectroscopy calculations and leptonic decay constant determinations.

 In QCD,  the effective running charge  and running quark masses, though not directly observable,  can be extracted from a suitable set of  experimental observables. At large momenta, QCD is asymptotically free \cite{GW1973,POL1973} as a consequence of the effective running coupling decreasing with increasing momentum transfer. At low momentum, the perturbative definitions suffer from unphysical singularities, or Landau poles. Hence, as rewieved in \cite{alfy} nonperturbative methods are required to calculate running couplings and the effective charge of QCD in this region.
 Alternatively, the perturbatively calculated running charges are  smoothed based on a set of analytical assumptions \cite{NE2003,BNPS2008}. 
  The useful concept  of the  effective charge have been recognized
in Bethe-Salpeter studies of mesons \cite{BRW1994,AGNEPA2007,HPGK2015,func2018,NAHP2021} as well as  appreciated in the  in its 3-dimensional  instantaneous approximation \cite{BNPS2008}.
The semi-perturbative method of the Pinch Technique allows to define the QCD effective charge for  effectively massive gluon in a gauge invariant manner \cite{CO1982}.
The effective QCD charge is then further moderated by the presence of the gluonic mass scale $m_g\simeq 1 GeV$.

In this paper, we examine the Bethe-Salpeter equation (BSE) for vector charmonia and investigate the impact of the running quark mass on the mass spectrum and leptonic decays. 
The  findings indicate that this effect is intriguing and the amount of fine-tuning is required. Together with preferred explanatory simplicity, this prevents us from achieving the ideal case in which we could solve the Schwinger-Dyson equation for the charm quark propagator, substitute it into the BSE, and examine physical outcomes such as the mass spectrum, decays, and other possible observables. This approach has already been applied to heavy quarkonia with moderate success
\cite{saulieta,saulipsy,sauli2025,HPGK2015,sauli2025} . 
However, when the running coupling is implemented more rigorously, we found that precisely determining leptonic decay constants requires an enormous accuracy and fine tuning of the BSE solutions. 
This problem does not arise in the nonrelativistic limit and may be avoidable through common three-dimensional reductions. In the present study, we aim to avoid the consequences of three-dimensional BSE reductions, the consequences of which are not yet fully understood. Instead, we maintain the full four-dimensional covariant setup, but simplify it.

The quantity under consideration in this study is the heavy quark propagator, expressed as follows:
\be  \label{run}
S(p)=\not p - <m(\xi)>\,  ;
\ee
where $<m(\xi)>$ is the mean effective charm  mass at the  scale $\xi$, which is linked to the  mean value  of the four-momentum $p^2$ 
 carried by the valence quark inside a given quarkonium state.   Recall that our definition is closely related to Grunber's   renormalization scheme invariant dynamical mass \cite{Grunberg}.
Taking   the scheme difference into account, the extracted values of the quark masses are found to be in alignment with the running masses that have been extracted by DSEs \cite{sauli2023} ans by alternative methods \cite{zeus2014,hera2017}. It is also helpful to  consider the instantaneous approximation of the BSE  for a moment. Then the scale  can be determined like  $\xi=<p^2>=<\bold{q^2}>+M_{\psi}^2/4$ where $M_{\psi}$ where
$p$ is the quark momentum,  $\bold{q}$ is the space  part of exchanged gluon momentum between the quark and antiquark. The estimates were made in \cite{BNPS2008}, 
giving  $<\bold{q^2}>=0.4 GeV^2$ for s-states charmonia. In our approach, the quarkonium state will be actually used to specify the scale.

\section{Quarkonia and Their Leptonic Decays in the BSE Formalism}

Charmonia are quarkonia composed of a charm  and an anti-charm  quark. Their masses differ significantly from those of other neutral mesons, thereby preventing mixing.
  Below the open charm threshold of $DD$ production, several narrow excited states exist, offering an ideal theoretical laboratory for our study. Here, we present the solution for the $1^{P=-1,C=-1}$ charmonium state, for which the most stringent constraints are imposed by experimental data on the leptonic decay constants.

The leptonic decay of all psi mesons can be calculated from a simple formula:
\be \label{botom}
F_{V} M_{V}=\frac{1}{3} Tr_{CD}\int\frac{d^4q}{(2\pi)^4} [\gamma_{\mu} S_c(q_+)\Gamma_V^{\mu}(q,Q) S_c(q_-) ]  \, ,
\ee
where $S_c$ stands for charm quark propagator and $q_{\pm}=q\pm Q/2$. 
The formula (\ref{botom}), as  proposed in \cite{IKR1999} and first used  in \cite{MATA1999}, 
assumes  that the meson pole part does not enter in QED Ward identity,  $\Gamma_V^{\mu}$ is thus a fully  transverse Bethe-Salpeter vertex
 function satisfying  $\Gamma . Q=0$.

The transverse function $\Gamma_V$ is called the  bound state vertex function and  satisfies the homogeneous BSE
\be \label{BSE}
\Gamma(p,Q)=i\int\frac{d^4k}{(2\pi)^4} S(k_+)\Gamma(k,Q)S(k_-)V(k,p,Q) \, ,
\ee
where $Q$ is the momentum of the meson with the mass $M$, satisfying $Q^2=M^2$.
The function $V(k,p,Q)$ in the Eq. (\ref{BSE}) is the  quark-antiquark scattering (ireducibile) kernel,  which is approximated by the product of the dressed quark-gluon vertex and the gluon propagator.  Additionally, we assume a gauge inavriant  Schwinger mechanism for generating the mass scale associated with gluons as suggested by V. Mathieu an collaborators in  \cite{AIMP2012}.
This is achieved by implementing the following choice of BSE interaction kernel:
\be \label{kernel}
V(k,p)=\gamma^{\nu}_T(q)\times \gamma_{\nu} 2 \alpha(q)\left[\frac{3/2}{q^2-m_g^2}
-\frac{1}{q^2-\Lambda_c^2}\right] \, ,
\ee
where $q=p-k$, $\gamma_T$ stands for  dirac matrices transversely  projected with respect to gluon momentum $q$,   
 $m_g\simeq 1.5 GeV$ is the effective gluon mass.  
This is  in a good agreement with other studies , e.g.  with  \cite{CO1982}.
$\Lambda_c$ was found to be an useful damping mass parameter;$\Lambda_c^2\simeq 25 GeV^2 $, and $\alpha$ is 
an infrared smooth    QCD running coupling, for wwhich we use the following formula
\be
\alpha(q)=\frac{4 C\pi}{(11-2N_f/3)ln(K-\frac{q^2}{\Lambda^2})}      \, ,
\ee
where we take $N_f=3$ active flavors, the constant $\Lambda$ is the perturbative QCD scale, numerically $\Lambda^2=0.1 GeV^2$ was taken. We use two methods of BSE solutions and provide the  parameters the Section below.  For the constant $C=1$, the prefactor in Eq. (\ref{kernel}) ensures the perturbative one loop UV asymptotic of the  running coupling  is chosen here.  However, for practical reasons both parameters $C$ and $K$ have been varied in order to get optimized charmonium spectrum and leptonic decay constants.

Relevant approximations can be summarized here:

i)The spin structure of the quark-gluon vertex is limited to the $\gamma^{\mu}$ and  the higher irreducibile kernels, with more complicated 
momentum structure, are not taken into account. 

ii) The running coupling is taken in a separable form, $\alpha((k+p)^2)\rightarrow \alpha(k^2+p^2)$, which allows integration of the analytical angular part of the kernel in the same manner as for the constant coupling approximation \cite{sauli2025}.

iii) The higher excited states that decay into $DD$ pairs are known to be  highly affected by open flavor channels. In this case, coupling with other channels have been ignored and we calculate their masses and leptonic decays using an idealized approximation.

Another  crucial assumption is that the observed charmonia are normal bound states and  not the ghosts. The solutions must satisfy  the canonical normalization 
condition (see for instance \cite{IKR1999}) and we exclude  parameter space that yield ghost solutions from our consideration. 

Lastly, we should mention that we do not use any additional phenomenological interactions. In other words, we avoid using a covariant version of the confining potential.  While introducing it is straightforward, and the author has used it earlier \cite{saulieta,saulipsy}, its presence  does not seem to be necessary for determining the spectrum of excited states and their  leptonic decay constants.  Of course, having more parameters in a game makes it easier to fit the theory to experiment. However, we do not use it because its presence would compete with the similar effect of the running quark mass, complicating the analysis.

\section{Solution of BSE and the quark mass determination}

This section describes the method for extracting the effective quark mass. To accomplish this, , we need to solve the BSE and properly normalize it. 
 We solved the BSE using two different methods. Variational methods have been used for the quarkonium system as early as the 1980s, primarily to study the dependence of energy levels on quark mass, as discussed in  \cite{DHP1981}. 
 These methods are employed in  contemporary light-front Hamiltonian approaches \cite{Choi2007,RAM2025}. 
 We used the variational method but also  solved the two-component BSE  (acommodated from \cite{sauli2025}) with the eigenvalue method. Here we remind the  choice of two componets which complete the iv-the point of approximations
described in the previsous sestion. 

iv)  Of the eight BS components for vectors, we use the following two:
\be \label{comp}
\Gamma^{\mu}(q,Q)=[\gamma^{\mu}- \frac{\not Q Q^{\mu}}{M^2}] \Gamma_1(q,Q)+ 
q^{\mu}\Gamma_5(q,Q) \, .
\ee

\begin{figure}[t]
\centering
\begin{minipage}{.5\columnwidth}
  \centering
  \includegraphics[width=7.4cm]{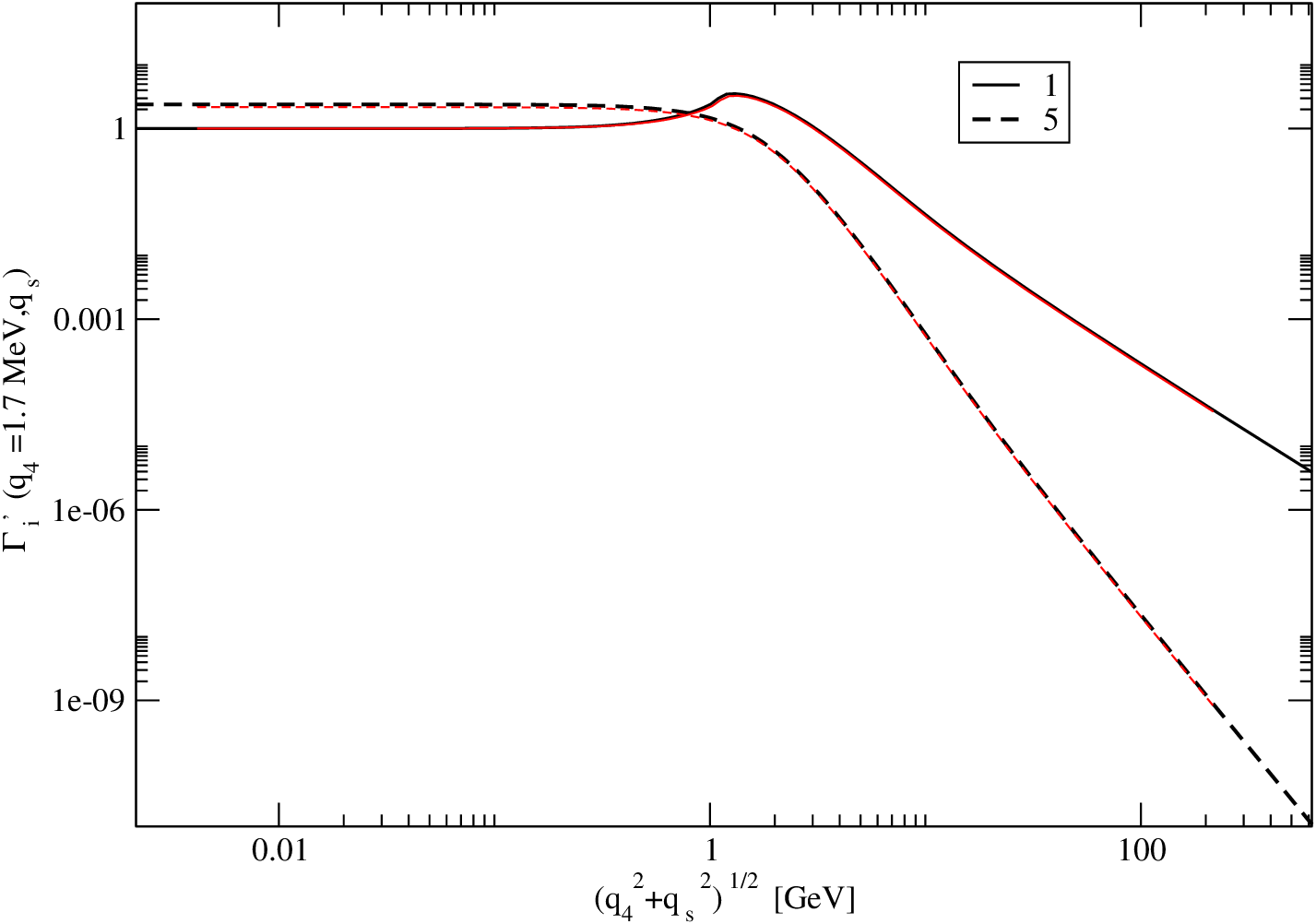}
  \caption{The first and the fifth component of $J/\Psi$ BS vertex functions as described in the text.}
  \label{wavesy1}
\end{minipage}%
\begin{minipage}{.5\columnwidth}
  \centering
  \includegraphics[width=7.4cm]{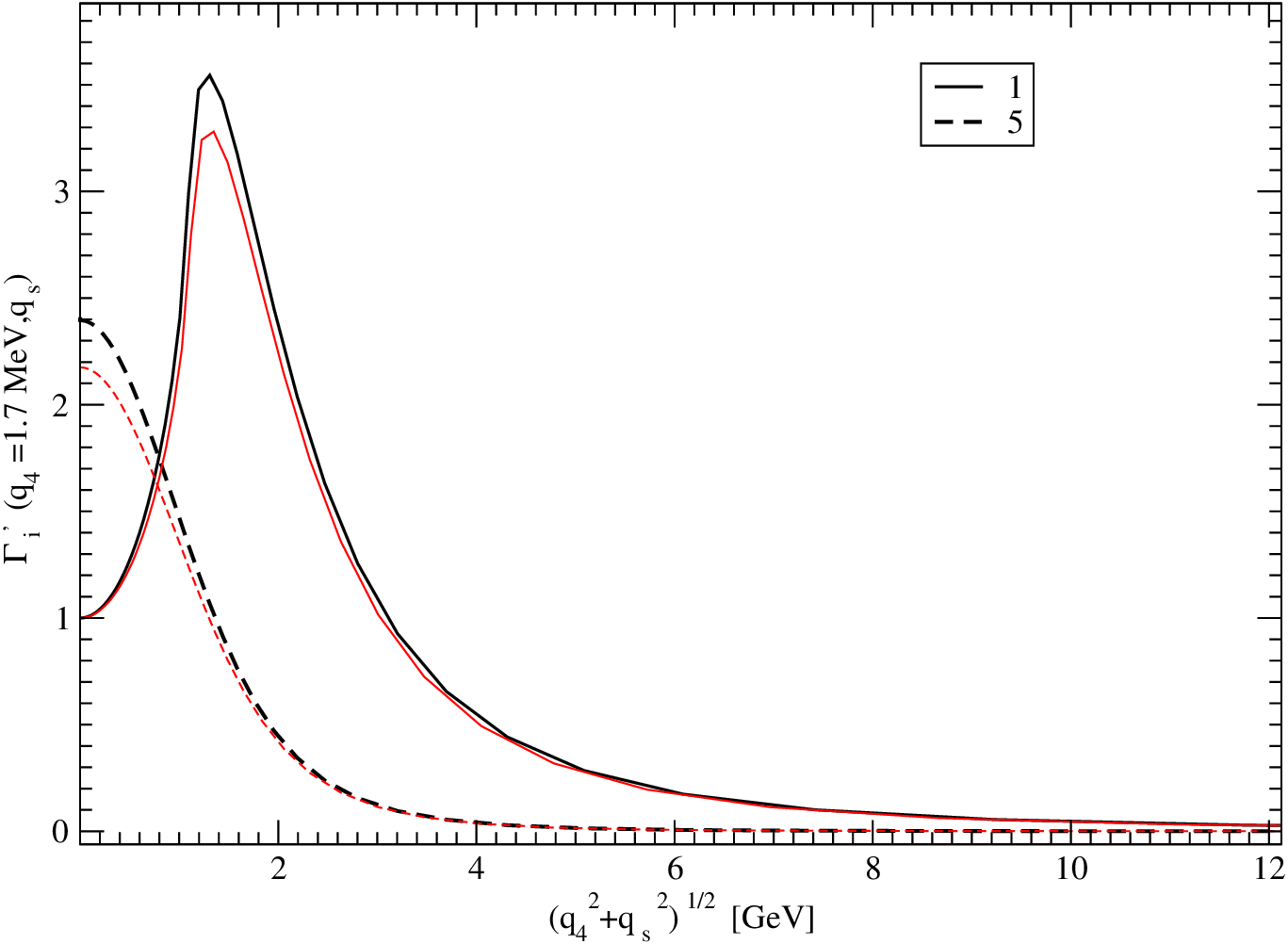}
  \caption{The same but in linear scale.}
  \label{wavesy2}
\end{minipage}
\end{figure}

 We set the gluon mass scales to be 
$m_g^2=2 GeV^2$ and $\Lambda_c^2=25 GeV^2$, respectively. The parameter $K=8 $ and the effective masses $<m_c>$ were 
determined using the  variational method described below.  These values were also used for the two-component case. 
Notably,  the two-component solutions with the running coupling differ radically from those obtained with the constant coupling approximation \cite{sauli2025}.
 The asymptotic behavior   is clearly visible  in the Fig. \ref{wavesy1}, where we plot the  $J/\psi$ BS vertex against the square of the relative (Euclidean) momentum $q^2$.  

Since the eigenvalue method is time-consuming, it is advantageous to start with the variational method.
To prepare the BSE for variations  we perform Wick rotation and  transform the BSE
into the a two-dimensional equation by performing angular integrations exactly as described in  \cite{sauli2025}. 
Then, we substitute  the trial Bethe-Salpeter vertex function $\Gamma^{trial}$ into the BSE and integrate it. To achieve the desired solution, we iterate and look for the optimal solution  by  minimizing the  normalized difference:
\be \label{chicko}
\chi_V^2= \frac{\sum   (\Gamma^{trial}(k,Q)-\Gamma^{l.h.s}(k,Q))^2 }{\sum   (\Gamma^{trial}(k,Q)+\Gamma^{l.h.s}(k,Q))^2 }.
\ee 
where  $\Gamma_{l.h.s}(k,Q)$ is the result of the integration. In Eq. (\ref{chicko}) the  sums run over the continuous  four-momentum $k$, discrete indices as well as over the BSE components.
The BSE solution obviously coincides with the vanishing limit $\chi^2\leftarrow 0$. We use the single component $\Gamma_1$ approximation to make the variational method more tractable.

To find the quark mass  means   determining  the propagator $S$ in Eq. (\ref{run}), which we actually use in our BSE calculation.
We determine such $S$ by minimizing $\chi_V$ , while also  checking that it   provides the correct leptonic decay constant. 
To this end, we  evaluate the covariant normalization condition and determine $F_V$ using a  normalized solution.

For the purposes of the  presented study, we  relax with $\chi_V^2\simeq  few 10^{-3}$, which is sufficient
enough to determine the effective quark mass using the combined knowledge of the meson mass and the decay constant. 
Due to the high precision required  for calculating leptonic decays, we chose the variational solution rather than the one obtained by the eigenvalue method.
A suitable choice of trial functions is essential and is described in detail in the Appendix.

We present the results in two tables. The first lists the narrow states. The second table is for charmonia above open charm thresholds, where the use of homogeneous BSE becomes unreliable. Since we do not couple to the decay channels, the agreement is not expected above the $DD$ thresholds. We added the second excited state to both tables because its mass passed only the first such threshold. 

It is important to note that the decay constant $f_V$ depends intriguingly on the quark mass $<m_q>$, allowing its numerical value to be quoted with a few percent accuracy.
When the interaction kernel is fixed, the values of the effective charm masses are unique and distinct for each quarkonium state.  
The observed scale  dependence appears to be a rigid feature of  the BSE solution for charmonia.  
The results for narrow states are summarized in Tab \ref{tabrho}. The quoted values are solely due to the experimental error of the decay constant $f_V$.

\begin{figure}[t]
\centering
\begin{minipage}{.5\columnwidth}
  \centering
  \includegraphics[width=7.4cm]{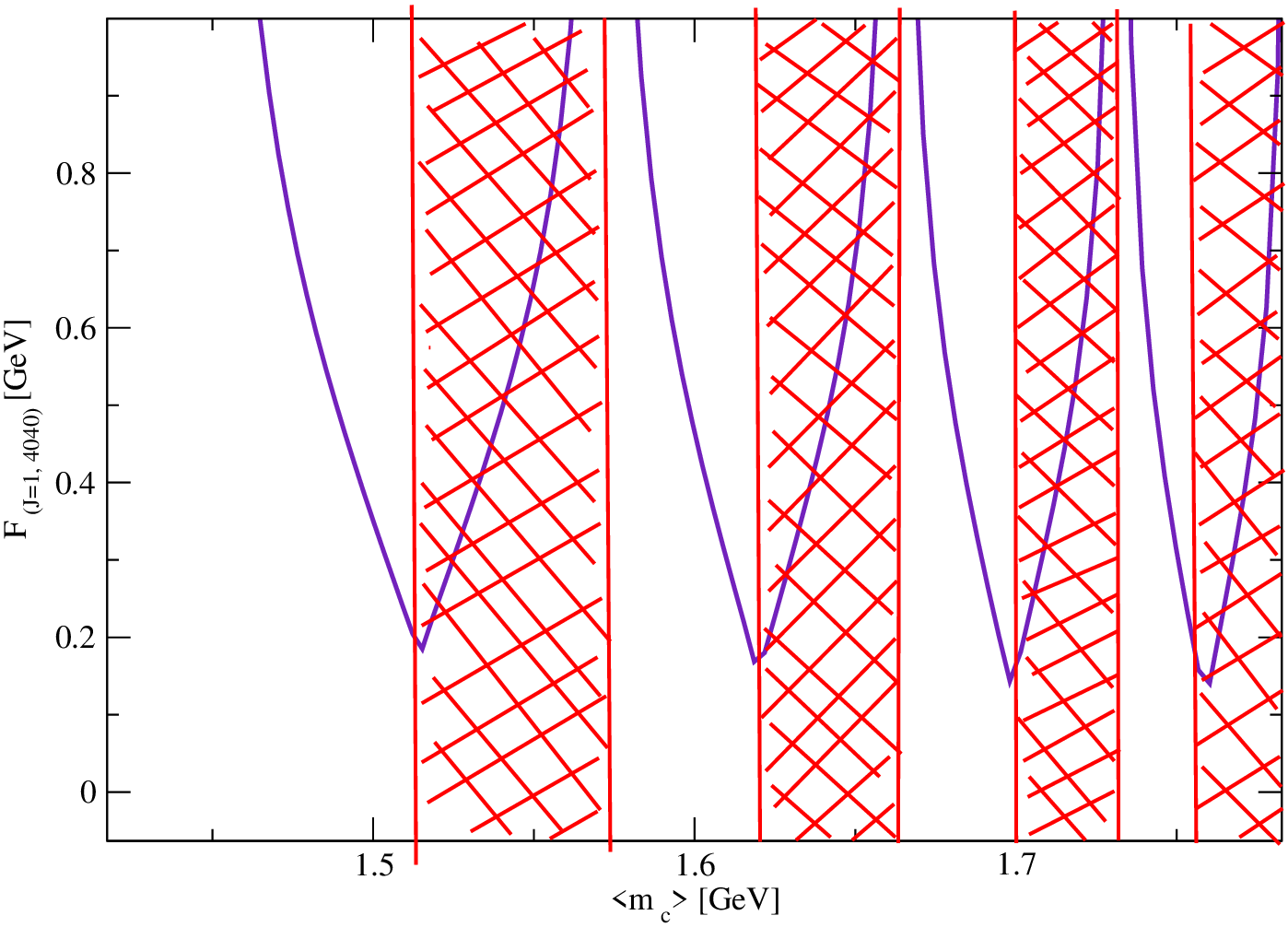}
  \caption{$F_{3s}$ against quark mass. The hatched area is exluded.The visible lines in excluded area correspond to unphysical, purely imaginary $f_V$,}
  \label{dva}
\end{minipage}%
\begin{minipage}{.5\columnwidth}
  \centering
  \includegraphics[width=7.5cm]{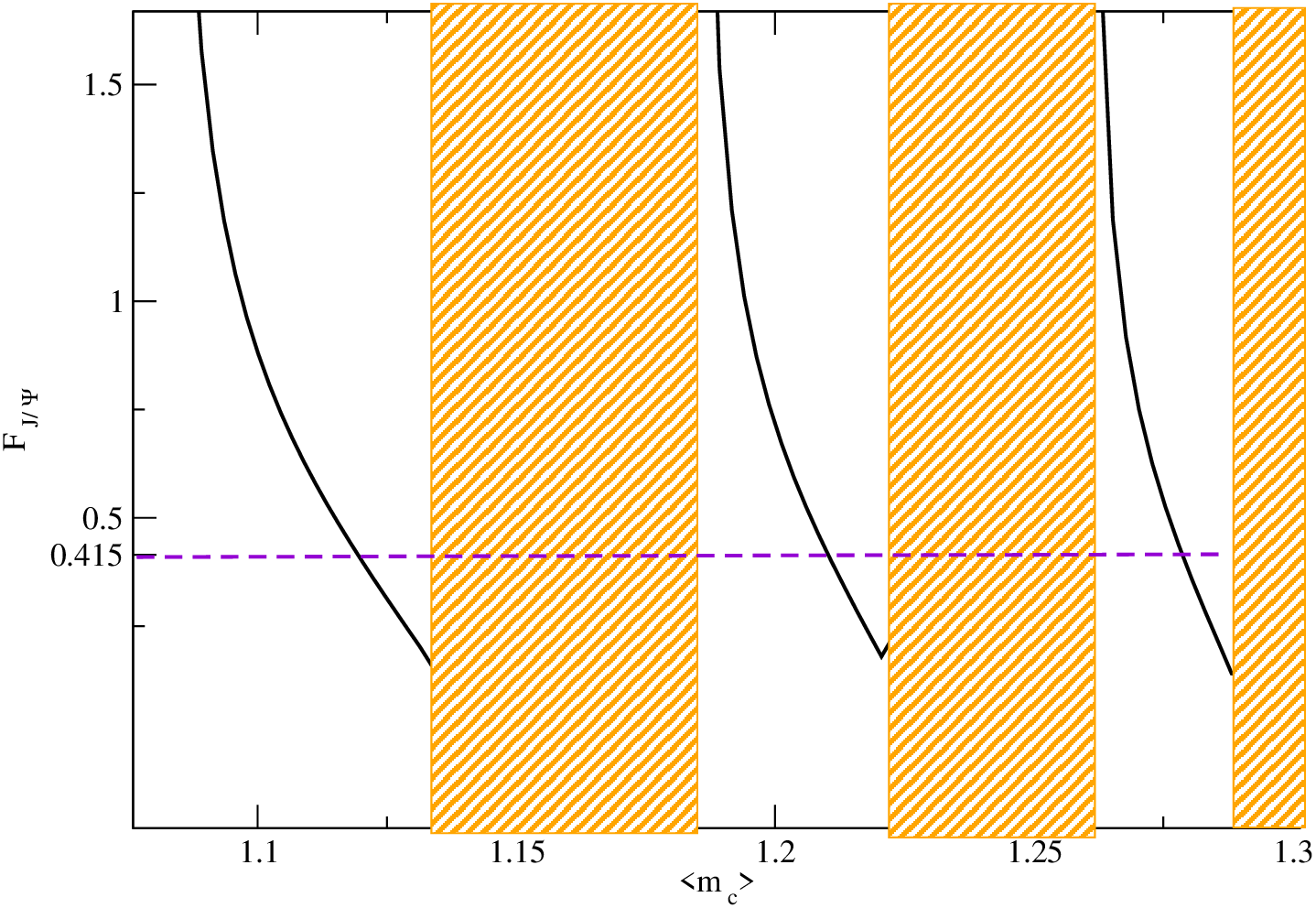}
  \caption{Variation of leptonic decay constant for $J/\psi$ shown in physical area only.}
  \label{tri}
\end{minipage}
\end{figure}

The  plot of  the function $f_V$ against the effective quark mass
is shown in the figures \ref{dva} and \ref{tri} respectively. Note, however, that the figure is merely illustrative and not exact since the function $f_V(<m_q>)$ is only precise near the variationaly determined value of $<m_q>$.

\begin{center} 
\begin{table}
\begin{tabular}{ |c|c|c|c| }
\hline
 name(mass)               &  $ f_V [MeV] $   &  $<m_c(M_V/2)>  $ & $\chi^2 $     \\
\hline
 $ J/\Psi(3100)$          &  415(4) &  1117(8)        & $5. 10^{-3}$                   \\
$ \Psi(3600) $            & 287(3)  &  1308(6)       &  $7. 10^{-3}$                                     \\
$\Psi(4040)$              & 238(5)  &   1514(10)     & $5. 10^{-3}$             \\       
 \hline
\end{tabular}
\caption{\label{tabrho} Used meson masses used (in parentheses) and experimental leptonic decays and extracted effective mass in MeV. 
Normalized deviation $\chi^2$ is shown in the last collumn. }
\end{table}
\end{center}

The heaviest charmonium with yet measured $f_v$ is labeled $\psi(4040)$ and listed in the Tab $\ref{tabrho}$ and its mass  lies already above $DD$ threshold.
In addition we use the same BSE as for narrow states but solve it  above $DD$ threshold. Rather than fitting to the experimental value, we chose the effective quark masses so that the hypothetical leptonic decay constants of open charm charmonia remain close to 200 MeV. These results are listed in Tab. \ref{tabrho2}.

The results in the first tab are fixed by comparison with the experiment, whereas there is more freedom in the case of an open charm meson. For this reason, the $4100$ state listed in the Tab \ref{tabrho2} is, in fact, the shifted $4040$ state of the Tab. \ref{tabrho} exhibiting thus the size of the systematic error.
And we estimate that there is a systematic error of 60 MeV due to ignoring the broad resonance character of the charmonia above the open-charm threshold.

The observed universal feature is the existence of ghost solutions outside normal state windows, creating prohibited gaps in solutions. The heavier the state, the more limiting the known leptonic decay constant measurements become.   An interesting question is whether this feature will survive beyond our effective mean-value quark mass approximation, Eq. \ref{run}.

   A notable challenge in the functional formalism of the Dyson-Schwinger and Bethe-Salpeter equations is typically addressed through numerical methods. Consequently, we address the associated numerical errors.   
The variational method was implemented by utilizing the experimental data to fit the solution. The numerical codes were formulated in such a manner that the masses were designated as inputs. However, in order to obtain the desired leptonic decay constants, additional effort was necessary. Errors resulting from such a search may be arbitrarily reduced. The systematic errors that have been achieved are presented in brackets at each table for the sake of comprehensibility.
 In the course of solving the BSE by means of the eigenvalue method, the masses were confirmed with a precision of $50$ MeV for each predetermined variational solutions obtained. 
A minor discrepancy may be attributable to the comparison of single (variational) and two (eigenvalue) components approximation; however, according to the findings of this study, the primary discrepancy is attributable to the weakness of the iterative/eigenvalue method.
The convergence of the eigenvalue method does not align with the ideal case observed in toy BSE models employed for pedagogical purposes.
 To identify the numerical error, the quark masses were varied and the solution was sought. 
In reality, it is difficult to reach the   intended value   $\lambda=1$ of the eigenvalue. The method employed in the recent stage provides no better resolution then $50$ MeV any meson. 
The numerics performs better for the higher states then for the ground state, i.e. for $J\Psi$ in our case. In the subsequent particular case, the tabeled quark mass $<m>=1.115 GeV$ provides  $\lambda=1.03$ and the $J/\psi$ mass  $M_{J/\Psi}=3161MeV$, while when taking  $<m>=1.114 GeV$ we get  $M_{J\Psi}=3120MeV$ but  with worse eigenvalue $\lambda=1.14$. 
All other model parameters were identical to the variational case, in which the experimental  mass $M_{J/\Psi}=3100MeV$ was established  by construction.  Not yet normalized BS vertex functions are plotted for the purpose of generating the figures \ref{wavesy1} and \ref{wavesy2} respectively. The dimensionless function  $\Gamma_5/1 GeV$ is actually shown. For purpose of completeness, the eigenvalue function actually used for the BSE solution in presented study is written in the Appendix.

\begin{center} 
\begin{table}
\begin{tabular}{ |c|c|c|c| }
\hline
 name(mass)               &  $ f_V [MeV]$   &  $<m_c(M_V/2)>  $ & $\chi^2 $     \\
\hline                              
$\Psi(4100)$              & 238(5)  &   $ 1578\pm 20 $  & $3. 10^{-3}$             \\
$ \Psi(4300)$             & 200(*)   & $ 1573\pm 20 $    & $ 4. 10^{-3} $  \\        
$ \Psi(4600)$             &  230(*) & $ 1750 \pm 20$ & $5. 10^{-3}$        \\
$ \Psi(4900)$              &  230(*) & $ 1750 \pm 20$ & $4. 10^{-3}$  \\     
 \hline
\end{tabular}
\caption{\label{tabrho2}Meson masses obtained in the BSE model and $f_V$ inputs by hand for heavier states in a 
hypotetical world without presence of light quarks.
We name the states according to the value of obtained masses by BSE solution. 
Note, $\Psi(4100)$ is a true minimum variant of $\Psi(4040)$, not a new state. 
Vector of the mass $4040$ in the previous Tab should be  understood as constrained approximation of broad resonance seen 
in the experiment. The stars indicate $f_V$ choosen by the author, note that  for  heavier states then 4300 MeV the solution for $f_V$ such that $f_V<230 MeV$
do not exist.}
\end{table}
\end{center}

\section{Conclusion and further prospect}

    We solved the BSE for excited vector charmonia using two methods. In addition to the standard eigenvalue technique, the minimization/variational technique was primarily useful for our purpose.
 Along with known masses, we used experimentally observed leptonic decays to determine the net structure of the effective quark mass at each quarkonium mass scale.
Using the kernel approximation based on the effective QCD running coupling, we obtained a covariant solution for the BSE vertex function provided that the effective quark mass varies with the scale.

Given that we do not use a confining or Wilsonian potential as part of the irreducible interaction kernel 
and given the experimental errors on the spectrum and the decay constant, 
the effective quark mass is then uniquely determined at a given meson scale. 
In our simple QCD model, we found that the decay constants and mass spectrum could not be reproduced when using a single constituent quark mass value.
In a near future, we plan to go beyond our approximation by considering the running quark masses from the theory. 
We also plan to investigate what would be amount of confining Wilsonian loop interaction required when electromagnetic properties of mesons are considered.

\appendix
\section {Trial functions for the BS bound state vertex}  
 We use the  leading component function $\Gamma_1(q,Q)$ (see the Eq. \ref{comp}) 
to obtain the solution variationaly.
This scalar function depends on two continuous variables: square of the  relative four-momentum $q^2$, and the product $q.Q$. Alternatively, in given frame, one can choose $q_4$ and $q_s$ for convenience.  Note that  
$q.Q/M=q_4$ in the rest frame of the meson, where the BSE is conventionally  solved.

After gaining some experience, the trial function used in this paper was chosen in the following form:
\be
\Gamma(q,P)=\sum_{r,s=1}^{N_r,N_s} C_{rs} \frac{d_{s}-o_r}{\left[(d_s-o_r)^2+c_s o_r^2\right]}
\frac{1+c}{\left[1+c ln^{3/2}(e+4 q^2/Q^2)\right]} \, ,
\label{trial}
 \ee
where $d_s=q_E^2/3-M^2/4-z_s(iq.Q)^2$ with Euclidean momentum notation for the relative momentum $q_E; q_E^2>0$, while the total momentum  $Q$ remains in Minkowski space, such that the meson has its physical mass $Q^2=M^2$. Note that the vertex function is a real-valued function  in the rest quarkonium frame.

The coefficient $C_{rs}$ is the weight that indicates the importance of a given fractional function that we use in the expansion of the trial functions.
It depends on two variables, $o_r$ and $z_s$, that we searche for by minimizing (\ref{chicko}) with suspicious intervals.
  In practice, we start with the nonrelativistic unbounded approximation, $z_s=0$, to identify the most important coefficients.  After fitting, we consider the new pair $z_1$ and $o_1$ and search for the triple $C_{1,1}, z_1, o_1$ by minimization again.
Once the search is complete, a novel (reduced) minimization of $\chi$  is performed
with the old parameters.  Another triple, $C,Z,O$, is added, and the procedure is repeated.  At each step, we reduce the suspicious interval for the variables included in the previous rounds.  While this does not guarantee the best fit with a given set of variables, the search is fast enough to reduce the chi-squared value to an acceptable size. Usually, when $N_{r}(N_s)$ equals to two or three it offers a satisfactory small $\chi^2$ of  a few $10^{-3}$ is achieved.   

 The   presence of a logarithm in Eq. (\ref{trial}) was found to be particularly useful  for all studied states.

For example, we review the parameters as achieved to establish the vertex function for the second excited charmonium state with a mass $M=4040MeV$.
Dimensionful quantities are in $GeV$. 
\be
  C_{r s} =
  \left[ {\begin{array}{cccc}
    0.37 & 2.11 & 0.47 &-1.05 \\
    -0.117 & -1.75 & 0.1 & 0.471 \\
  \end{array} } \right]
\ee

Where for  $s=1,2$ we use $d_1=M^2/4+q_E^2/3+0.415(M q_4/8)^2 $, $c_1=0.387$ and $d_2=M^2/4+q_E^2/3+(M q_4/4)^2 $, $c_2=0.382$.
The vector $o$ was fitted such that (in units of GeV) $o=(0,2.26,3.62,4.9)$. 
For the log prefactor we get  $c=0.225$ and the variables $z_s$ are explicitly shown in polynomials $d_s$. 

A similar number of parameters was used to fit the vertex functions for all states, 
including the  two narrow states $J\Psi$ and $\Psi(2)$.  
The nonrelativistic  wave functions for ground states  are generally quite simple: they have no nodes.   As expected, a simplification occurs in the BS treatment as well.
It turns out that the vertex for the $J/\psi$ meson can be chosen  independently of   the invariant $q.Q$ and it satisfies 
a simple form:   
\be 
\Gamma_{J/\psi}^{trial}(q)=\frac{d-o}{\left[(d-o)^2+c_1 o^2\right]}
\left[ln(c+\frac{q^2}{M^2/4})\right]^{-\gamma}
\ee
with $d=q^2+K P^2$ it achieves  relatively small $\chi^2=4.8 \, 10^{-2}$ with the five  parameters in need: 
$o=2 GeV^2, K=2 ; c_1=0.435; c=7.5; \gamma=3/4 $. Pushing $\chi^2$ downwards
requires more parameters,  similarly to the case of $3s$ state  described above.

\section{The BSE eigenvalue function}

 We have used the  eigenvalue function which is identical to the numerator of the  difference $\chi_E$ used to quantify  the quality of  the iteration procedure.             
\bea \label{anocho}
\chi_E^2&=& \frac{\sum   (\Gamma^{i+1}(k,Q)-\Gamma^{i}(k,Q))^2 }{\lambda^2(Q)} \, ;
\nn \\
\lambda(Q)&=&\sqrt{\sum(\Gamma^{i+1}(k,Q)+\Gamma^{i}(k,Q))^2 } \, ;
\eea 
 where $i$ is the iteration step number.  The sum signifies the double integrations with the suitable weight function $W$
impleneted. In our case we evaluate in the BSE in the rest of the meson, where $Q=(M,0)$, a useful choice, that we have used reads 
\bea
\lambda(Q)^2&=&\int_{-\infty}^{\infty} dk_4 \int_0^{\infty} dk_s W(k,P)  (\Gamma^{i+1}(k,Q)+\Gamma^{i}(k,Q))^2  \;
\nn \\
W(Q)&=&\left[Q^2/4+k_e^2\right]^{-1} \, ;
\eea
where $k_s$ is the magnitude of the space part of the Euclidean relative quark four-momentum $k_e$, i.e. $k_e^2=k_4^2+k_s^2$.

  A successful solution is characterized by Eq. $\chi_E^2\simeq 0$ and  trivial factorization, since $\lambda=1$. 
Bacause the integral equations for the BS components are mutually coupled, a single component can be used to evaluate
 the criterion  Eq. \ref{anocho}. 


%
\end{document}